\documentclass{iau} 
\usepackage{graphicx}

\title[Massive stars, successes and challenges] 
{Massive stars, successes and challenges}

\author[Georges Meynet et al.]   
{Georges Meynet$^1$, Andr\'e Maeder$^1$,
Cyril Georgy$^1$, Sylvia Ekstr{\"o}m$^1$, Patrick Eggenberger$^1$, Fabio Barblan$^1$, \and Han Feng Song$^{2,3}$}

\affiliation{$^1$Department of Astronomy, University of Geneva, Switzerland\\
$^2$College of Science, Guizhou University, Guiyang, 550025, Guizhou Province, PR China\\
$^3$Key Laboratory for the Structure and Evolution of Celestial Objects, Chinese Academy of Sciences, 650011, Kunming, PR China
 \\ email: {\tt georges.meynet@unige.ch}
}

\pubyear{2017}
\volume{xxx}  
\jname{Title of your IAU Symposium}
\editors{A.C. Editor, B.D. Editor \& C.E. Editor, eds.}
\begin{document}

\maketitle

\begin{abstract}
We give a brief overview of where we stand with respect to some old and new questions bearing on how massive stars evolve and end their lifetime.
We focus on the following key points that are further discussed by other contributions during this conference: convection, mass losses, rotation, magnetic field and multiplicity. 
For purpose of clarity, each of these processes are discussed on its own but we have to keep in mind that they are all interacting between them offering a large variety of outputs,
some of them still to be discovered. 
\keywords{Massive stars, convection, mass loss, rotation, magnetic field, multiplicity}
\end{abstract}

\firstsection 
\section{Massive stars in the Universe}
Stars are at the crossroad of many topical questions in astrophysics, cosmology and physics. 
In astrophysics, the way they form, evolve and end their nuclear lifetimes has a large impact on the evolution of the matter as a whole in the Universe, on the evolution of galaxies. Also stars are  bridges for connecting processes occurring at different space and time scales. For instance, stars are important producers of dust grains in the Universe (e.g. Todini \& Ferrara 2001) and at the same time are the sources of light and of new elements in galaxies (e.g. Chiappini et al. 2006). They also provide
links between evolution of stars at various redshifts and metallicities covering the whole cosmic history, from the origin of the very peculiar composition of the most iron poor stars (Frebel \& Norris 2015) to the origin of the short lived radionuclides in the nascent solar system (Gounelle 2015).
At a cosmological level, stars are sources of ionizing photons. This is especially true at low metallicities and for the first stellar generations that may have contributed significantly to the reionization of the Universe (see e.g. Amor\`in et al. 2017). 
Stars offer powerful physics laboratories (e.g. in the case of the solar neutrinos) to explore some questions at the frontiers of physics as the
possible variations with time of the fundamental constants (Ekstr{\"o}m et al. 2010), or properties of new particles as Weakly Interactive Particles or axions (see e.g. Taoso et al. 2008). 
The ranges of temperature and density conditions they span vastly outrange the domains that may be explored in the laboratory. Of course, to address all these questions, a first prerequisite is to sufficiently well understand their physics.
In the following, we discuss a few selected questions giving the opportunity to present some new ideas.

\section{Convection}
Convection, and more generally turbulence, is a long standing problem in stellar evolution. Actually these processes are difficult to describe already on Earth and thus it is no surprise that indeed they are difficult to be well described in stars. These processes involve many different space and time scales, they are fundamentally 3 dimensional processes and thus their incorporation in 1D models is done at the expense of some simplifications. 

Then why not to address this question through 3D hydrodynamical modelling? Actually 3D modelling is indeed used to probe the physics of convection, but this approach is possible only for relatively short time intervals. If one wants to explore the evolution of stars along their whole lifetime, exploring different initial masses, chemical compositions, rotation, magnetic field,... then 1D models are still the only tool that allows exploring large areas of the parameter space. This is the reason why some authors (Meakin \& Arnett 2007ab; Arnett et al. 2015; Arnett \& Meakin 2016) follow the approach consisting in deducing from multi D hydrodynamical simulations more physical recipes for incorporation into 1 D models. This is by far not an easy task. Actually it consists as written by Arnett \& Meakin (2016) to  
extrapolate the weather to determine the climate! Said in other words, it is not straightforward to deduce from short time simulations, prescriptions for convection that can be used on long term evolution. 

Stellar models usually use either the Schwarzschild or the Ledoux criterion for determining the size of the convective core. In addition some overshooting can be applied extending the size of the convective cores. During the Main-Sequence phase of massive stars, there is no changes expected whether the Schwarzschild or the Ledoux criterion is used. Indeed, in massive stars, during the Main-Sequence phase, the convective core decreases in mass, thus at the border of this receding core there is no $\mu$-gradient ($\mu$ being the mean molecular weight). But during the core He-burning phase where the core increases in mass, depending which criterion is used lead to different outputs. 

Gabriel et al. (2014) discuss various numerical methods to determine the size of the convective core in 1D models. During the iterative process for computing the structure of the star at a new time step, some numerical method has to be chosen in order to determine the size of the convective core at the new iteration. The boundary of the core is physically given by the condition $L_{\rm rad}=L_{\rm total}$, where $L_{\rm rad}$ is the luminosity due to the radiative transport and $L_{\rm total}$, the total luminosity also accounting for the thermal and kinetic energy transported by convection. In the local Mixing Length Theory  and without overshooting, this condition is equivalent to  the equality of the temperature gradients at the border of the convective core, {\it i.e. }$\nabla_{\rm rad}=\nabla_{\rm ad}$ {\it on the convective side of the boundary} (Biermann 1932). This condition on the gradients may not be reached on the radiative side of the boundary if a $\mu$-gradient is present in the radiative layers above the convective core.

At the end of a given iteration, we shall have values for the adiabatic and radiative gradients at every mesh points. The quantity $\Delta\nabla=\nabla_{\rm rad}$-$\nabla_{\rm ad}$ changes sign at the border of the core (passes from a positive value inside the convective core to a negative one in the radiative zone in case no semiconvective zone is present). To determine the position where $\nabla_{\rm rad}=\nabla_{\rm ad}$, there are three possibilities: 1) take the two last positive values of $\Delta\nabla$ at the border of the convective core and extrapolate to find the place where $\Delta\nabla$ becomes zero; 2) take the last positive value inside the convective core and the first negative one in the radiative one and interpolate to find the place where $\Delta\nabla$ becomes zero; 3) to extrapolate (inwards) the first two negative values of $\Delta\nabla$ in the radiative region. 

Gabriel et al. (2014) first shows that depending on that choice, various values for the convective core can be obtained. As a numerical example, they show that the mass of the convective core in a 16 M$_\odot$ during the MS phase (when the mass fraction of hydrogen in the core is about 0.15) is 22\% of the total mass when the rule 1 is applied and only 17\% when the rule 3 is applied (here adopting the Ledoux criterion). This produces a difference of more than 20\%! 
 A second important result that they have obtained is that the only physically justified method is the method number 1, that means interpolating from inside the convective core. Any other choice can lead to unphysical results (1) in case a convective core is expanding and/or (2) in case the Ledoux criterion is adopted. 

In case of a shell convective zone, at the moment no definite conclusion can be reached about the proper numerical procedure since in case of a convective shell forming in a mu-gradient region, it is impossible to satisfy the condition $L_{\rm rad}=L_{\rm total}$ at both boundaries. Actually, depending on which criterion for convection is chosen (although in the case of a pre-existing mu-gradient region, the Ledoux criterion should be adopted), significantly different results are obtained (see {\it e.g.} Georgy et al. 2014). 

\section{Mass losses}
Mass loss by stellar winds is a key process in massive star evolution. It changes the evolutionary tracks in the HR diagram, the surface abundances and velocities. The wind contributes to the chemical enrichment in new elements of the interstellar medium, to the rate of injection of momentum and energy in the interstellar medium. It has also an impact on the flux of ionising photons. Mass loss has also an impact  
on the nature of the supernova event (if any) and on the properties of the stellar remnant (if any) (see {\it e.g.} the review by Smith 2014). 

For hot stars, the line driven winds theory provides a sophisticated theoretical frame that can account  for the amplitude of the mass losses by stellar winds (with uncertainties limited to a factor of a few) and their dependence with the metallicity (see e.g. the review by Puls et al. 2008).  Note however that already a factor 2 difference during the Main-Sequence phase may produce significant changes at solar or higher metallicities for the stars with masses larger than about 30-40 M$_\odot$ (Meynet et al. 1994). Greater uncertainties are present when the stars
evolve away from the MS phase and enter into the regime of Luminous Blue Variable or of the red supergiant stage. In both cases, observations tell us that these stars may show outbursts. The LBV iconic case is $\eta$ Car that in the middle of the eighteen century lost during one or two decades mass at a rate of about 1 M$_\odot$ per year (see e.g. Humphreys \& Martin 2012). Less extreme, but still showing some sporadic mass ejection events are for instance the star VY CMa, a bright and extended evolved cool stars (See Humphreys 2016 and the references therein). 

Outbursts are difficult to be accounted for in stellar evolution computations mainly because their physics is still to be unravelled. These outbursts may however have an important impact removing in short timescales large amounts of mass
(see e.g. Smith \& Owocki 2006).  When occurring just before the supernova, they might produce superluminous supernovae by converting part of the mechanical energy of the ejecta in radiations.

These outbursts make typically the modelling of the LBV and also of the red supergiant phase still uncertain. To illustrate this point, it is interesting to compare the post red supergiant evolution obtained with different
prescriptions for the red supergiant mass loss rate (Salasnich et al. 1999; Vanbeveren et al. 2007, Georgy 2012, Meynet et al. 2015). 
As is well known, removing mass during the red supergiant stage may make the star to evolve back to the blue side of the HR diagram. 
In general the bluewards evolution for stars with masses above about 15 M$_\odot$ begins when
the mass of the core becomes a fraction higher than about some limit that depends on the initial mass and is around 60-70\% the total mass of the star (Gianonne 1967). 
Increasing the mass loss has thus for effect to shorten the red supergiant lifetime of those stars that eventually evolve bluewards. 
It produces core collapse supernova progenitors  that are yellow, blue supergiants, even sometimes LBV's or Wolf-Rayet stars. An interesting result of this kind of evolution is that it might affect, if frequent enough, the distribution
of stars as a function of the luminosity along the red supergiant branch. If we look at this argument the other way around, it means that the observed luminosity distribution of the red supergiants provides some hints about the time averaged mass loss rate during that phase. Larger are the mass losses, larger will be the decrease as a function of luminosity of the number of red supergiants (see for instance the left panel of Fig. 5 in Meynet et al. 2015). This might be interesting
to measure from complete red supergiant samples the slope of the luminosity distribution function and to compare with predictions of population synthesis models based on various mass loss rates during the red supergiant phase.

 \section{Rotation}
Axial rotation is a very interesting feature of stars for many reasons: 
\begin{itemize}
\item The angular momentum content of a star on the ZAMS results from the star formation process.  
As is well known, during the formation process large amounts of angular momentum have to be removed from the collapsing cloud otherwise stars cannot be formed. How this happens depends on processes like for instance disk locking, disruption of the collapsing cloud in multiple systems. Haemmerl\'e et al. (in press) computed pre-MS evolutionary models with accretion and rotation. They concluded that during the phase between the formation of the small mass hydrostatic core of  0.5 M$_\odot$ and the arrival on the ZAMS of the star with its final mass, in order to avoid the star to reach the critical velocity a braking mechanism is needed. This mechanism has to be efficient enough to remove more than 2/3 of the angular momentum from the inner accretion disc. They also conclude that due to the weak efficiency of angular momentum transport by shear instability and meridional circulation during the accretion phase,
the internal rotation profiles of accreting stars reflect essentially the angular momentum accretion history.
\item During its nuclear lifetime, rotation can induce many changes in the observed properties of stars (see e.g. the review by Maeder \& Meynet 2012 and references therein). It changes the surface abundances as a results of the mixing processes induced by the same instabilities indicated above that transport angular momentum. It modifies the evolutionary tracks in the HR diagram making a star of a given mass more luminous than its non-rotating sibling. Rotation may activate a dynamo in convective regions and it might also activate one in differentially rotating ones. When the star is rotating sufficiently fast, typically with surface angular velocity larger than about 70-80\% the critical angular velocity, the star will be become oblate, it will show anisotropic stellar polar winds (more intense winds in the polar rather in the equatorial direction). At very high rotation, the rotational mixing may be so strong that the star will follow a homogeneous evolution.
The surface velocity, together with other observed properties represents thus an important pieces of information of stellar physics, either for single stars or stars in close binaries.
It varies as a results of the internal redistribution of the angular momentum by convection, shear, meridional currents, magnetic fields instabilities, also by
processes like stellar mass losses, tidal interactions with a companion, or mass accretion from a companion. 
\item At the end of its stellar lifetimes, rotation may change the consequences of the core collapse. In case the core rotates sufficiently fast, it may favor a luminous supernova explosion. It may also have an impact on the
rotational properties of the stellar remnant if any. It is however not obvious how the state of rotation in the presupernova structure is linked to the rotation rate of the neutron stars and black hole. Indeed the explosion mechanism itself and/or some braking mechanism operating in the early phases of the evolution of the new born neutron star may have a significant impact and thus hide the pre-explosion conditions.
In case the angular momentum would be conserved, present day stellar models predict in general too fast rotating neutron stars when compared to the rotation rate of young observed pulsars (Heger et al. 2005). 
If true it would indicate that massive star models with rotation still miss some angular momentum transport mechanism, a process that appears well confirmed for small mass stars (see Beck et al. 2012, Eggenberger et al. 2017 and references therein).
\end{itemize}
The physics of rotation is complex and involves turbulence, a feature that, as recalled above cannot be described easily in numerical simulation. In 1D models, the unknown aspects of turbulence are
accounted for in the choice of a few parameters. 
As an example, in the Geneva stellar evolution code, we use the theory proposed by Zahn (1992) who is based on the hypothesis that the star settles into a state of shellular rotation due to a strong horizontal turbulence.
In those models two parameters related to the shear turbulence have to be fixed. 
One  is the value taken for the critical Richardson number. This value governs the efficiency of the mixing by the  shear instability along the radial direction. A second one intervenes in the expression of the horizontal turbulence. 

The Richardson criterion comes from the fact that for the mixing to occur, the excess energy from differential rotation  has to be larger than the energy needed to overcome the gradient of density (see the textbook by Maeder 2009).
The excess of energy in the shear can be expressed by $1/4 \rho (\delta V)^2$, where $\rho$ is the density, and $\delta V$ the differential of velocity over a given distance in the radial direction. The energy needed to overcome the vertical density gradient can be written
$g \delta\rho \delta z$, where $g$ is the gravity, $\delta\rho$ the difference of density between the interior of the blob and the exterior, $\delta z$ the length scale over which the blob moves. 
The Richardson criterion tells that one has mixing when $g \delta\rho \delta z < {1\over 4} \rho (\delta V)^2 $, or $R_{\rm i}=g \delta\rho \delta z/\rho (\delta V)^2 < R_{\rm i,crit}={1\over 4}$. In this approach the critical Richardson number is 1/4. Some numerical simulations indicate that turbulence begins to appear already for a value of $R_{\rm i,crit}=1$ (Br{\"u}ggen \& Hillebrandt 2001). The choice of $R_{\rm i,crit}$ is thus confined between 1/4 and 1 and this is one of the important parameter that has to chosen. 

The Richardson criterion is actually the key expression to find the value of the diffusion coefficient due to shear in the radial direction. A diffusion coefficient can be expressed as $(1/3) \upsilon l$ where $\upsilon$ is a typical velocity and $l$ a typical size of the moving blobs. In a turbulent medium, the transport is dominated by the largest eddies, {\it i.e.} the largest $l$ values. The size of the eddies enter into the expression of $\delta\rho$ through the way the transport changes the temperature and molecular weight gradients (see Maeder 2009 and references therein). Thus looking for the size of the eddies that  satisfies the Richardson criterion, it is possible to deduce  the vertical diffusion coefficient. In case of secular shear, {\it i.e.} shear occurring on timescales that are long compared to the thermal diffusion timescale, it is in general always possible to find a size of the eddies that satisfies the Richardson criterion.
Note that it is also important to check that the largest eddies that satisfy the Richardson criterion are not too small to be damped by the viscous forces. 

The second parameter intervenes into the expression for the horizontal turbulence, {\it i.e.} the turbulence along an isobar. In the last Geneva grid at solar metallicity (Ekstr{\"o}m et al. 2012), the choice  of these two quantities was mainly driven by requiring that stars with initial masses between 9 and 15 M$_\odot$ at solar metallicity, presenting a surface velocity during the Main-Sequence phase compatible with the observed averaged velocities present nitrogen surface enrichments in agreement with the mean observed values. Note that this  way of doing is shared by other groups (Brott et al. 2001, Chieffi \& Limongi 2013) and thus the stellar models will predict similar values in this mass and velocity range at the solar metallicity. However, outside these ranges, different models may actually predict significantly different behaviors depending on the way rotation is accounted for.

In shellular rotating models, when no magnetic field is present, the mixing of the chemical elements is mainly due to shear instabilities, while the transport of the angular momentum is mainly driven by the meridional currents.
In models where a strong internal magnetic field is considered, as for instance due to the Tayler-Spruit mechanism (Spruit 1999, 2002), the mixing of the chemical elements is mainly driven by meridional currents.

The stellar models predict that the surface enrichments in products of the CNO burning,
observable for example by the N/H, or N/C ratios, increase with stellar mass M and rotation
velocities v, because mixing gets stronger. The enrichments also increase with the age t, since
more and more new nuclear products reach the stellar surface. The
enrichments are also stronger at lower Z, shear mixing being favored in more compact stars.
Tidal interactions, as will be seen below, may influence the mixing. Thus, the chemical enrichments are
a multivariate function (Maeder, 2009), e.g. (N/H)=f(M, $\Omega$, age, Z, B,  multiplicity,....).
This fact has for consequence that it is difficult from observed stars to isolate the specific correlation 
between the surface enrichments and rotation. When sufficient care has been taken for selecting the stars 
presenting properties allowing to isolate the effect of rotation (isolated stars of about the same initial mass, age and metallicity)
then a good agreement is found between rotating stellar models and the observations (Maeder et al. 2009, see also Przybilla et al. 2010, Martins et al. 2015).

\section{Magnetic fields}
\begin{figure*}[b]
\begin{center}
 \includegraphics[width=2.5in]{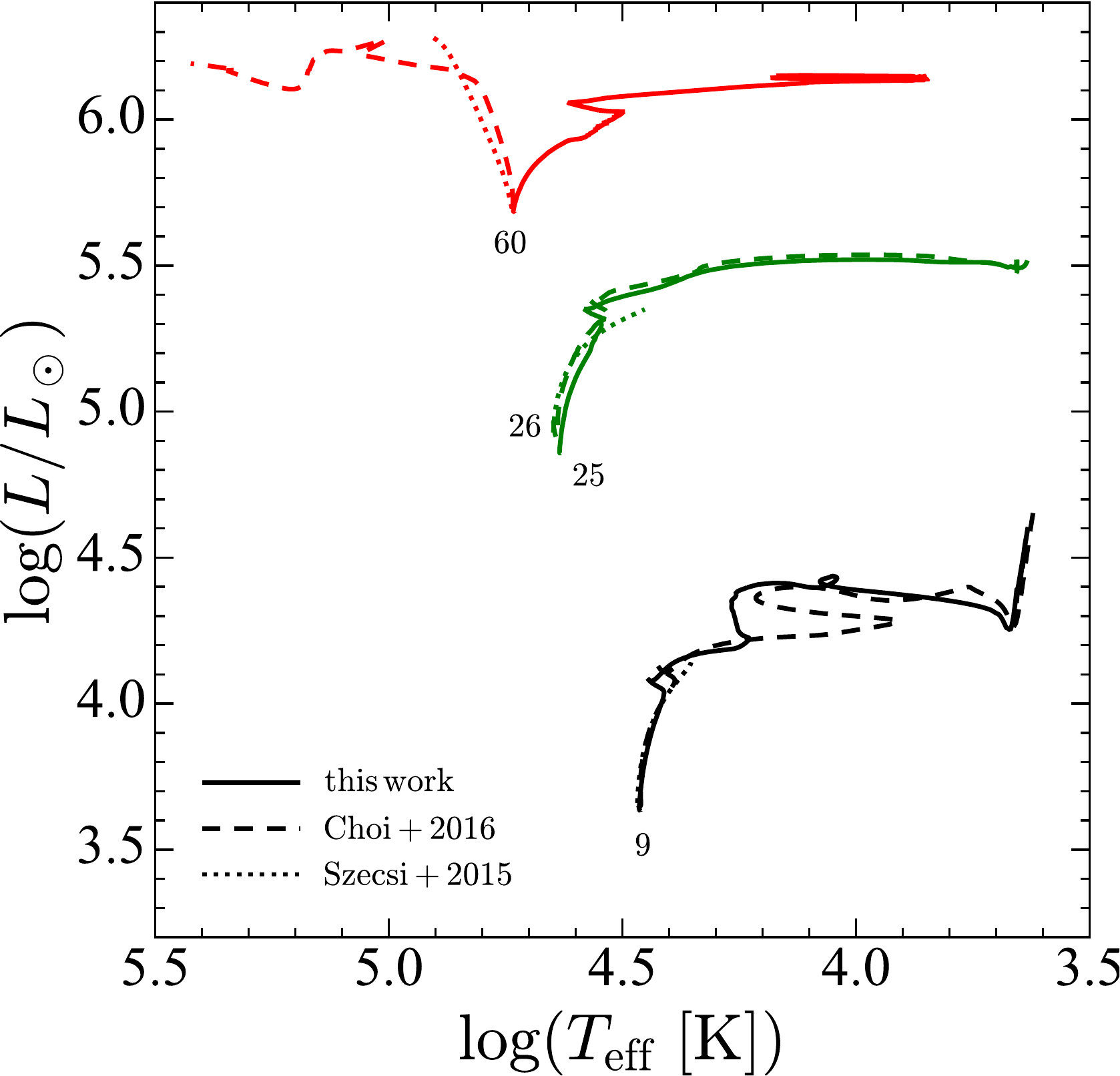} \includegraphics[width=2.5in]{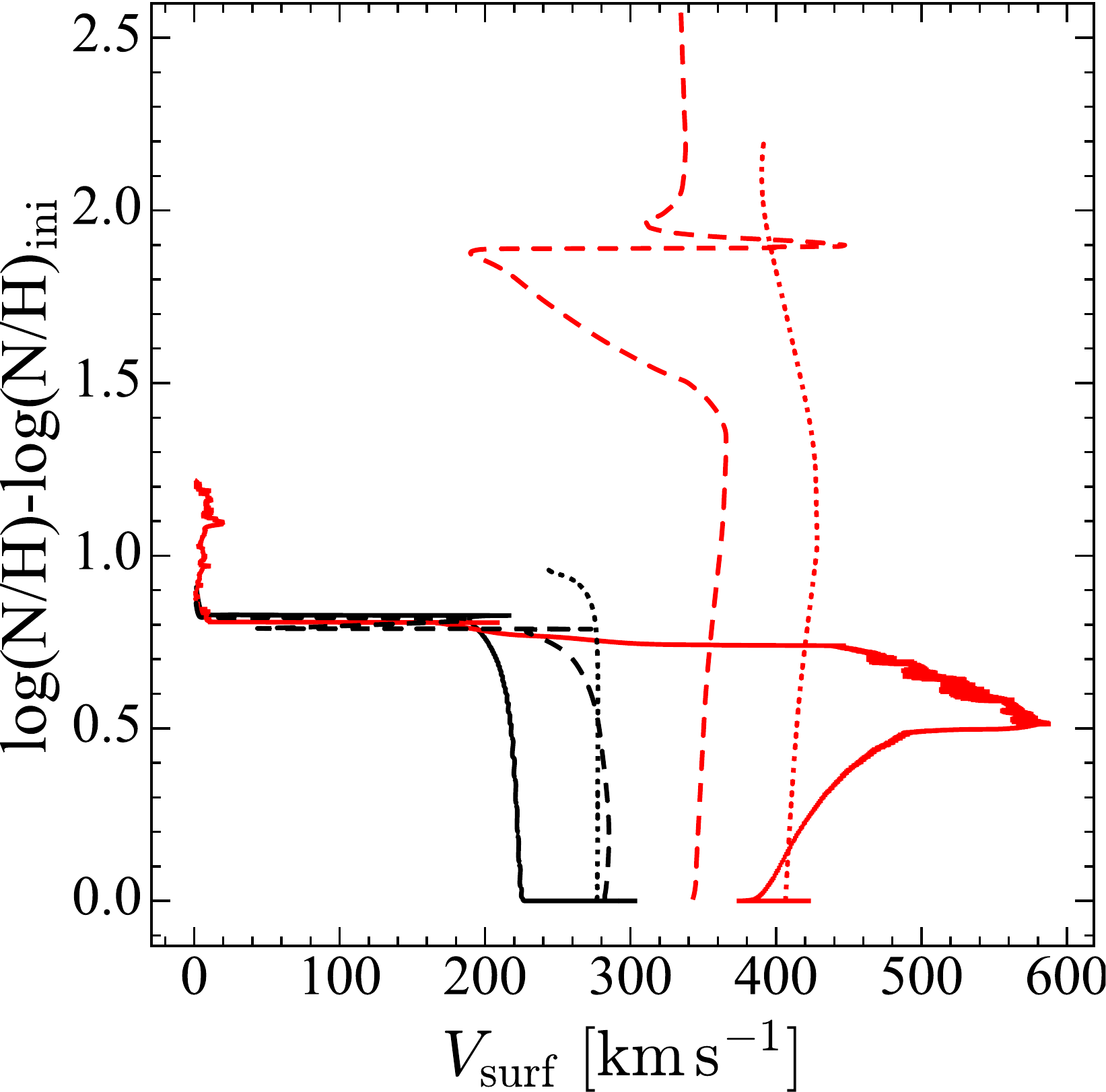}
 \caption{{\it Left panel:} Comparisons of the evolutionary tracks for different initial mass models in the theoretical HR diagram for a metallicity, $Z$ around 0.00004, {\it i.e.} corresponding to the metallicity of the galaxy I Zw 18. The code MESA without
an internal dynamo has been used by Choi et al (2016) and a [Fe/H]=-1.5. The code STERN with an internal dynamo has been used by Szecsi et al. (2015, note that they computed a 26 M$_\odot$ model). {\it Right panel:} comparison in the plane 
surface nitrogen abundance versus surface rotation. Only the 9 and 60 M$_\odot$ models are shown for purpose of clarity. Figure taken from Groh et al. (in preparation).
 }
   \label{figCOMP}
\end{center}
\end{figure*}


Magnetic fields can intervene in various parts of the star: starting from the core region, a magnetic field can be attached to the convective core. Then in the radiative envelope, a magnetic field can be amplified by the Tayler-Spruit dynamo (Spruit 1999, 2002). Finally at the surface, a magnetic field can be present (Wade et al. 2006; Grunhut et al. 2017), either produced by some dynamo attached to the small convective region that is present even in massive stars  in the outer layers (Cantiello et al. 2009), or more probably being a fossil field, {\it i.e.} having its origin in a previous phase of the evolution of the massive star. In that last case, it is reasonable to think that this magnetic field will actually be present in the whole star taking much larger values in the interior. 

A main impact of a surface magnetic field is to exert a torque at the surface of the star\footnote{A strong magnetic field may reduce the mass loss by stellar winds, having interesting consequences for the formation of massive black holes even at solar metallicity (Petit et al. 2017) or for Pair Instability Supernova to appear at solar metallicities (Georgy et al. 2017).}. This happens in case the energy density associated to the magnetic field is larger than the density of kinetic energy in the wind (ud-Doula \& Owocki 2002), so typically when
$B^2/(8\pi)$ is larger than $1/2 \dot{M} \upsilon_\infty^2$, where $B$ is the equatorial magnetic field, $\dot{M}$ the mass loss rate, and $ \upsilon_\infty$ the terminal wind velocity, {\it i.e.} the velocity of the wind when the process of acceleration is terminated. The impact of this wind magnetic braking is to slow down the star (ud-Doula et al. 2008, 2009). This slowing down may be accompanied or not by strong surface enrichments depending whether the magnetic field is due to a surface dynamo attached to the outer convective zone or if it results from a fossil magnetic field  present in the whole star (Meynet et al. 2011).

The impact of a strong magnetic field in the interior of the star is mainly to impose a nearly rigid rotation. This is the case for instance when the Tayler-Spruit mechanism is used to compute stellar models. Although this process will be active only in differentially rotating layers, it will actually make the whole star to rotate as a solid body because convective zones are assumed to rotate as solid bodies.
As already indicated above, in those models, the mixing of the elements is not long driven by shear instabilities since there is no strong differential rotation, but by meridional currents. A question that might be asked at that point is whether meridional currents will not be prevented to be active by the magnetic field itself! Likely this depends on the geometry and strength of the field (Zahn 2011). 

In Fig.~\ref{figCOMP}, the evolutionary tracks for different initial mass models having similar initial rotations are plotted for purpose of comparison. The metallicity of these models corresponds to that of the Galaxy I Zw 18 and is about 1/50 the solar metallicity. One of the main differences between the Geneva tracks and the tracks computed with STERN is that in the Geneva tracks, the Tayler-Spruit dynamo in radiative zones is not accounted for while this mechanism is taken into account in the other code. The most striking difference occurs for the 60 M$_\odot$ models. While the Geneva track, for an initial velocity of about 430 km s$^{-1}$ evolves as usual to the red part of the HR diagram, the STERN model with magnetic field, starting with an initial rotation of 340 km s$^{-1}$ evolve homogeneously. This is an effect mainly due to the different physics considered. Actually the Geneva code with the account of the Tayler-Spruit dynamo would also produce a similar behavior.  Thus we see that the inclusion of the Tayler-Spruit dynamo favors the homogeneous evolution. The MESA code produces for the 60 M$_\odot$ model (410 km s$^{-1}$) a strong transport of angular momentum and of the chemical species even without an internal magnetic field.
This comparison shows that it is important to be aware that behind the terms {\it rotating models}, very different physics may be considered leading to significantly different outputs. 

If we compare the changes of the surface abundances predicted by these two types of models (see the right panel of Fig.~\ref{figCOMP}), we see that indeed the MESA and STERN tracks are much more enriched than the Geneva track, quite consistently with the behavior in the HR diagram. We can note another difference in this right panel. The MESA and STERN tracks evolves from the beginning vertically, while the Geneva track evolves first horizontally (the surface velocity is decreasing) and then evolves vertically. The initial decrease of the velocity occurs on a very short timescale and is due to an initial redistribution of angular momentum by the meridional currents inside the star. This redistribution is triggered by meridional currents that transport angular momentum from the outer layers to more central region, hence the decrease of the surface velocity.

\section{Multiplicity}
\begin{figure}[b]
\begin{center}
 \includegraphics[width=5.0in]{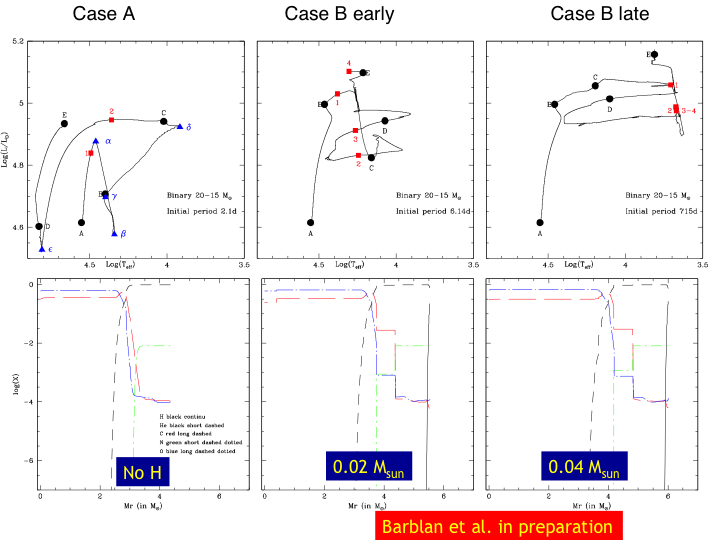} 
 \caption{The three upper panels show the evolution of a non-rotating 20 M$_\odot$ at solar metallicity in a binary system with a 15 M$_\odot$. The initial orbital period is respectively from left to right equal to 2.10, 6.14 and 715 days. In each panel, point A corresponds to the ZAMS, B to the end of the core H-burning phase, C to the ignition of helium in the core, D to the end of the core He-burning phase, and E to the end of the core carbon burning phase.   A first mass transfer episode occurs between point 1 and 2. A second mass transfer occurs between points 3 and 4. The lower panel shows the chemical structure of the stars at point E. }
   \label{figBIN}
\end{center}
\end{figure}

\begin{figure}[b]
\begin{center}
 \includegraphics[width=2.2in]{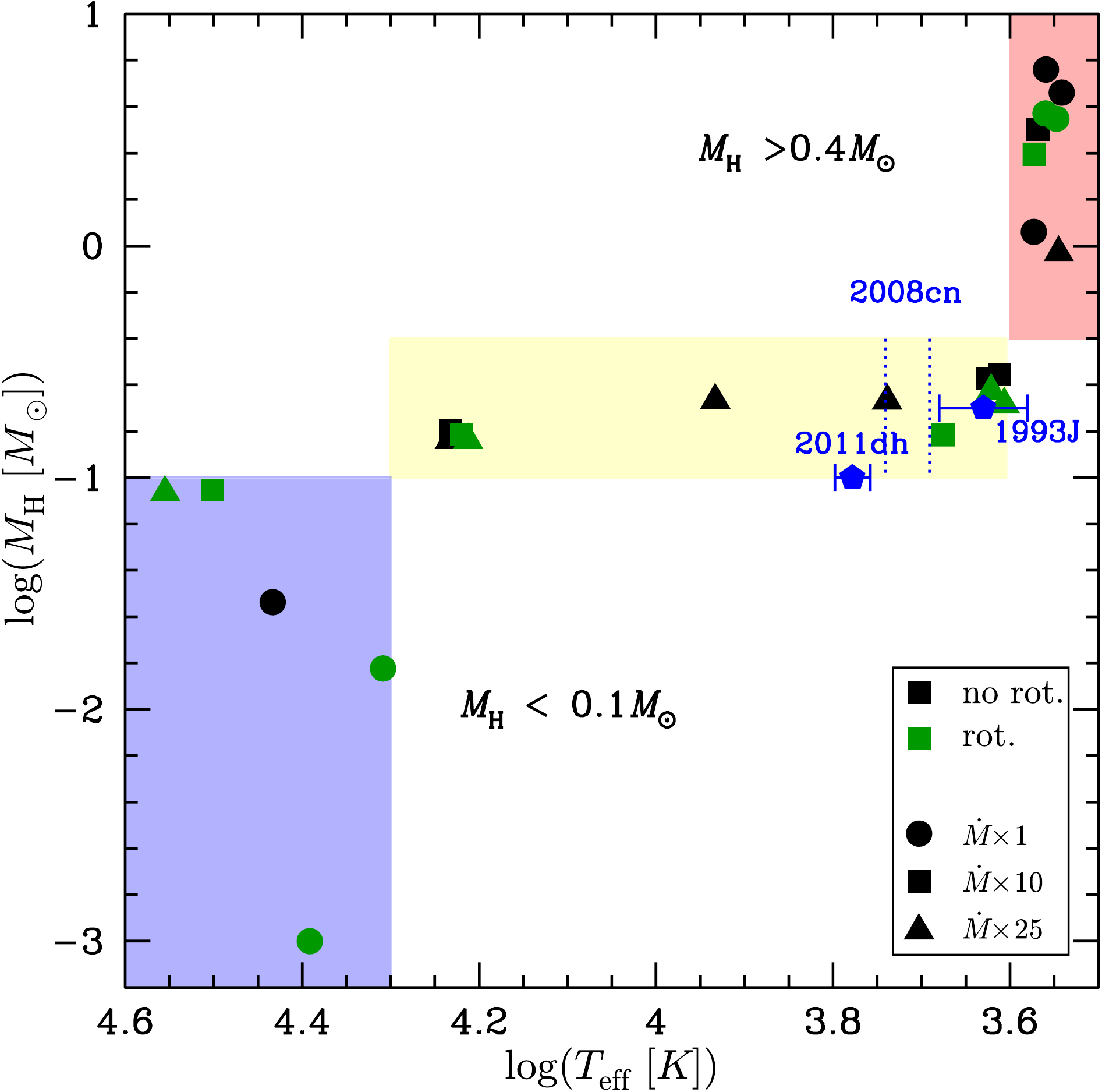}  \includegraphics[width=3.0in]{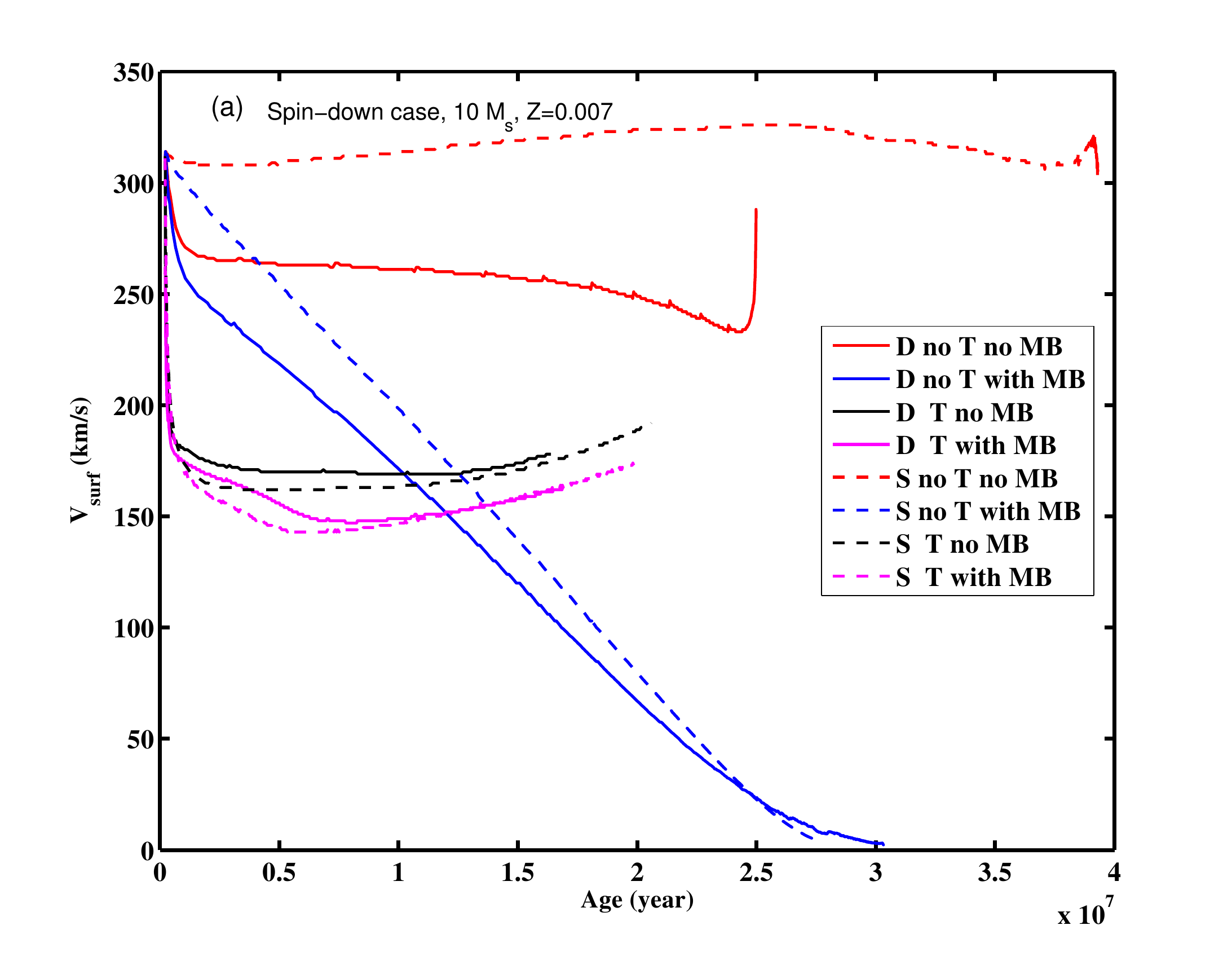}
 \caption{{\it Left panel:} Mass of hydrogen in solar masses at the pre-supernova stage for the various models with initial masses between 9 and 25 M$_\odot$ at solar metallicity, for various initial rotation between 0 and 40\% of the critical velocity on the ZAMS and for different prescriptions of the mass loss rates during the red supergiant phase (see more detail in Meynet et al. 2014). Positions in this diagram of some supernovae are indicated by pentagons with error bars. 
The big red dots
correspond to the positions in that diagram of the three 20 M$_\odot$ binary models shown in Fig.~\ref{figBIN}.
{\it right panel}: Evolution of the surface equatorial velocity as a function of time for different 10 M$_\odot$ stellar models at a metallicity $Z$=0.007 with $\upsilon_{\rm ini}=310$ km s$^{-1}$. The symbols D means (radially) differentially rotating models,
S, solid body rotating models, T, models with tides in a close binary system (the companion is a 7 M$_\odot$ star and the initial orbital period is 1.2 days), MB is for wind magnetic braking (the equatorial surface magnetic field is 1 kG).} 
\label{figHTEFF}
\end{center}
\end{figure}

Many massive stars are in multiple systems and part of those may follow a different evolution because of interactions with their close companions (Sana et al. 2012, 2013).  
This has triggered many recent works exploring close binary evolution and their consequences for explaining the origin of various stellar populations (see e.g. Eldridge \& Stanway 2009, 2016; Stanway et al. 2016; Yoon et al. 2010, 2017).
Examples of the evolution of a primary star of 20 M$_\odot$ 
in a short period binary systems with a 15 M$_\odot$ are shown in Fig.~\ref{figBIN} (non-rotating stellar models). Three different cases are shown, corresponding to different initial orbital periods and thus different times for the first mass transfer episode. These models were computed in order to see whether the primary could evolve into a low luminous WC star at the end of the evolution. Such low luminous WC stars are observed (Sander et al. 2012) and thus the question is how they are formed. Do they result from the single star channel? In that case it would required very high mass loss rates. It might be the case if some outbursts occurred during the evolution of the progenitors as can be reflected by the fact that many WR stars present ring nebulae (see e.g. Esteban et al. 2016). Do they results from close binary evolution? To provide at least a partial answer to that last question we computed the models shown in Fig.~\ref{figBIN}. We chose for the primary a 20 M$_\odot$ model because its luminosity, in case it would evolve into the WC stage would
more or less match the luminosities observed for the low luminous WC stars. Actually, the primary will evolve into WNE stars but it will never reach the WC phase. Thus this channel does not provide a solution for the origin of these stars. It has then to be checked whether the secondary might evolve into that stage. Another question is why, in case such an evolutionary channel would be frequent enough, the WNE stars that it produces are not observed. A possibility is that
these stars may be difficult to detect hidden in the light of their more massive companion having accreted the mass. This point needs also to be confirmed by more detailed investigations.

In the following, we shall use the models of Fig.~\ref{figBIN} to illustrate another effect of binarity.  Looking at the bottom panels of Fig.~\ref{figBIN}, we can see the chemical structures of the different models at the core carbon-ignition. The outer layers at that stage have already reached their final structures, unless very strong mass loss episode would still occur in the very last moments of the evolution just before the supernova explosion. A feature that does appear as a kind of special features that is seen in these close binary models is the tiny amounts of hydrogen that is left in the cases B mass transfer (early and late). Actually single star evolution may have difficulties in producing such structures. This is illustrated in the left panel Fig.~\ref{figHTEFF} where the mass of hydrogen in the envelope of pre-supernovae models is given for various models as a function of the effective temperature. All the models, except the large red starry dots are from single rotating and non rotating stellar models computed with various mass loss rates during the red supergiant phase. The models with a mass of hydrogen larger than 0.4 M$_\odot$ explode as red supergiants, and all the models with masses of hydrogen below about 0.1 M$_\odot$ are Wolf-Rayet stars just before the core collapse. Only those models that have a mass of hydrogen intermediate between these two values are yellow-blue supergiants. If we plot on that same diagram the models obtained by close binary evolution,
two of them fall in the blue region as other models obtained by single star evolution. However there is the interesting case of the late case B mass transfer that falls below the yellow shaded region, indicating that this channel is able to produce a core collapse supernova progenitor with a lower hydrogen content at a given effective temperature. Interestingly this late case B would more or less mimic an increase  of the mass loss rate during the red supergiant stage, however it does not give a similar structure as the models with an enhanced red supergiant mass loss. This is due to the fact that the single and the close binary models have different mechanism for putting an end to the strong mass loss episode. In the case of the single star, what makes the star to evolve into a lower mass loss rate regime is simply the evolution out of the red supergiant stage when a critical amount of mass has been lost. In the case of the close binary, it is the the decrease of the primary radius below the critical Roche limit that puts an end to the mass transfer.  Thus increasing the mass loss rate at the red supergiant stage will not necessarily produce the same structure as a mass transfer episode occurring at the red supergiant phase. 
One can wonder, whether more generally a low hydrogen content (at least in this mass domain between 15 and 25 M$_\odot$) might be an indication favoring a core collapse supernova progenitors having gone through a mass transfer episode. Much more computations need to be performed. If true it might be an interesting signature since the mass of hydrogen in the progenitor can be sometimes trace back from properties of the supernova light curve.

The physics of rotation is important to model close binary stars. The reason is that in a binary there is a huge reservoir of angular momentum in the orbital movement. Through tidal interactions, exchanges between the orbital and the axial orbital reservoirs happens, sometimes spinning up the stars, sometimes spinning them down and of course changing the parameters of the orbit. The impact of these changes of angular momentum in stars and the way this angular momentum is redistributed inside the star by 
various instabilities is important for questions regarding many aspects of the evolution of such systems like synchronization, circularization, induced tidal mixing etc...
Thus rotation and binarity might be tightly intertwined. 

Also rotation, multiplicity and magnetic winds can interact. Recently Song et al. (in preparation) have investigated the case of massive stars with a strong surface magnetic field in a close binary system. The question these authors want to address is how tidal and magnetic torques interact and what are the consequences for the axial rotation of the two stars and for the orbital evolution. The right panel of Figure~\ref{figHTEFF} shows what happens in a system composed of a 10 and a 7 M$_\odot$ star orbiting around their center of mass with an orbital period of 1.2 days. Cases with and without a surface magnetic field for the 10 M$_\odot$ has been considered (the equatorial surface magnetic field considered is taken equal to 1kG)
The plot shows the evolution of the surface velocity with time of the primary. Two different angular momentum distribution has been considered, solid body rotation driven by the Tayler-Spruit dynamo and the case of differential internal rotation driven by shear and meridional currents. The cases of single stars with and without magnetic braking are also shown. We see that the single stars when braked reach very low surface velocities at the end of the Main-sequence phase . The same star with a companion will actually be maintained at a much higher rotation by the tidal torques. Tidal torques tap angular momentum from the orbit to transfer it to the star. This tends to reduce the distance between the two stars and thus to increase the tidal torque. This is why the velocity of the primary increases after 5-10 Myr. We see that
the difference between the solid body and differentially rotating case is the most important in the case of the single non-magnetic stars. As soon as either the wind magnetic braking or the tidal torque (or both) occur, the two cases show very similar behaviors in the surface velocity versus time plane. For what concerns the evolution of the tracks in the HR diagram and the surface abundances, the differences are much larger.



\section{Conclusion}
The few questions addressed above do not pay credit to many other interesting questions that massive star evolution still trigger. One sees that the picture becomes more complicated adding to the impact of the initial mass and metallicities, other quantities as the initial rotation, the magnetic field and the multiplicity. We saw that these effects can interact strongly: for instance rotation has an impact on the lifetimes and the evolutionary tracks, this in turn changes the quantities of mass and angular momentum lost by the stellar winds, changing the rotation of the star. In a close binary system, rotation may change under the impact of mass loss and by tidal torques. These processes have an impact on the orbital evolution and thus
on the tidal torques.

Likely the challenges for the future will be on one side to obtain the correct physics to account in a proper way of all these effects in stellar models. But even if that stage will be reached, then will remain the challenge of exploring the consequences of many different initial conditions and to find a reasonable way to compare with the observations. While these challenges are severe, there is some hope that they will be at least in part overcome in the future thanks to the ever increasing observational channels that provide new constraints about the way stars are evolving. The recent detection by LIGO of gravitational wave well illustrates this point.  Large surveys collecting data on huge number of stars and thus offering unbiased samples of observed stars are also essential for making progress. Even short lived phases may have very important consequences. Thus we are living in a very exciting time for exploring the physics of massive stars that are so important
for driving short timescales processes in our Universe.

\begin{acknowledgement}
GM thanks Prof Arlette Noels for her help and the hospitality of the Kavli Institute in Santa Barbara.
\end{acknowledgement}


\end{document}